\documentclass{emulateapj}
\usepackage{longtable}

\begin{document}

\title{Well-Sampled Far-Infrared Spectral Energy Distributions of \lowercase{z} $\sim$ 2 Galaxies: Evidence for Scaled up Cool Galaxies}

\author{Adam Muzzin\altaffilmark{1}, 
Pieter van Dokkum\altaffilmark{1},  Mariska Kriek\altaffilmark{2}, Ivo Labb\'{e}\altaffilmark{4}, Iara Cury\altaffilmark{1}, Danilo Marchesini\altaffilmark{3},  \& Marijn Franx\altaffilmark{5}}

\altaffiltext{1}{Department of Astronomy, Yale
  University, New Haven, CT, 06520-8101; adam.muzzin@yale.edu} 
\altaffiltext{2}{Department of Astrophysical Sciences, Princeton
  University, Princeton, NJ, 08544} 
\altaffiltext{3}{Department of Physics and Astronomy, Tufts University, Medford, MA 06520, USA}
\altaffiltext{4}{Hubble Fellow, Carnegie Observatories, 813 Santa
  Barbara Street, Pasadena, CA, 91101}
\altaffiltext{5}{Leiden Observatory, Leiden University, PO Box 9513,
  2300 RA Leiden, Netherlands} 
 
\begin{abstract}
We present an analysis of the far-infrared (FIR) spectral energy distributions (SEDs) of two massive K-selected galaxies at $z =$ 2.122 and $z =$ 2.024 detected at 24$\micron$, 70$\micron$, 160$\micron$ by Spitzer, 250$\micron$, 350$\micron$, 500$\micron$ by BLAST, and 870$\micron$ by APEX.  The large wavelength range of these observations and the availability of spectroscopic redshifts allow us to unambiguously identify the peak of the redshifted thermal emission from dust at $\sim$ 300$\micron$.   The SEDs of both galaxies are reasonably well fit by synthetic templates of local galaxies with L$_{IR}$ $\sim$ 10$^{11}$L$_{\odot}$ -- 10$^{12}$L$_{\odot}$ yet both galaxies have L$_{IR}$ $\sim$ 10$^{13}$L$_{\odot}$.  This suggests that these galaxies are not high redshift analogues of the Hyper-LIRGs/ULIRGs used in local templates, but are instead "scaled up" versions of local ULIRGs/LIRGs.  Several lines of evidence point to both galaxies hosting an AGN; however, the relatively cool best fit templates and the optical emission line ratios suggest the AGN is not the dominant source heating the dust.  For both galaxies the star formation rate determined from the best-fit FIR SEDs (SFR(L$_{IR}$)) agrees with the SFR determined from the dust corrected H$\alpha$ luminosity (SFR(H$\alpha$)) to within a factor of $\sim$ 2; however, when the SFR of these galaxies is estimated using only the observed 24$\micron$ flux and the standard luminosity-dependent template method (SFR(24$\micron$)), it systematically overestimates the SFR by as much as a factor of 6.  A larger sample of 24 K-selected galaxies at $z \sim$ 2.3 drawn from the Kriek et al. (2008) GNIRS sample shows the same trend between SFR($24\micron$) and SFR(H$\alpha$).  Using that sample we show that SFR($24\micron$) and SFR(H$\alpha$) are in better agreement when SFR($24\micron$) is estimated using the log average of local templates rather than selecting a single luminosity-dependent template, because this incorporates lower luminosity templates.  The better agreement between SFRs from lower luminosity templates suggests that the FIR SEDs of the BLAST-detected galaxies may be typical for massive star forming galaxies at $z \sim 2$, and that the majority are scaled up versions of lower luminosity local galaxies.  
\end{abstract}

\keywords{infrared: galaxies}

\section{Introduction}
It is well known that in the local universe the obscuration towards star forming regions is correlated with star formation rate (SFR; e.g., Wang \& Heckman 1996; Hopkins et al. 2001; Buat et al. 2005, 2007), with the most active galaxies frequently being the most dust obscured.  In the most extreme galaxies, the star forming regions can become optically-thick (e.g., Genzel et al. 1998), which makes mid- and far-infrared (MIR, FIR) data critical for a complete metric of their SFR.  At $z \sim$ 2, where the average massive galaxy is forming stars at rates $\sim$ 100 - 1000 times higher than in the local universe (e.g., P\'{e}rez-Gonz\'{a}lez et al. 2005; Juneau et al. 2005; Damen et al. 2009), MIR- and FIR-determined SFRs are critical for understanding the buildup of today's massive galaxies.  
\newline\indent
The advent of the MIPS instrument onboard $Spitzer$ has facilitated the first deep photometry at 24$\micron$ and numerous studies have used these data to infer the star formation history of the universe up to $z \sim$ 2 (e.g., Le Floc'h et al. 2005; P\'{e}rez-Gonz\'{a}lez et al. 2005; Caputi et al. 2007; Reddy et al. 2008; Magnelli et al. 2009).  These studies suggest that obscured star formation dominates in massive galaxies at $z \sim$ 2; however, the conversion from observed 24$\micron$ flux to a total infrared luminosity (L$_{IR}$), and subsequently a SFR at $z \sim$ 2 requires significant extrapolation based on local FIR templates.  
\newline\indent
The most common method for determining the L$_{IR}$ of a galaxy from 24$\micron$ photometry is to artificially redshift local luminosity-dependent MIR/FIR templates to the redshift of the galaxy and then choose the L$_{IR}$ of the template that predicts a 24$\micron$ flux closest to the observed 24$\micron$ flux.
By applying this method, or more sophisticated but related versions of it, some authors have found that the local templates produce estimates of L$_{IR}$ and SFR(L$_{IR}$) for distant galaxies that are in agreement with other SFR indicators (e.g., Elbaz et al. 2002, Marcillac et al. 2006, Reddy et al. 2006; Daddi et al. 2007a; Papovich et al. 2009; Kartaltepe et al. 2010), while others have argued that this method tends to systematically overestimate the L$_{IR}$ and hence the SFR (e.g., Papovich et al. 2007; Rigby et al. 2008; Murphy et al. 2009).  Some of these differences can probably be reconciled by the different selection criteria  (e.g., Reddy et al. 2010), redshift range (e.g., Murphy et al. 2009) and luminosities (e.g., Papovich et al. 2007) of the above samples; however, the reliability of observed 24$\micron$ fluxes as a metric of the L$_{IR}$ and SFR over the full range of galaxy masses and SFRs at $z \sim$ 2 remains unclear.  
\newline\indent
The next step will be to directly observe the cold dust SED, rather than extrapolate it from the rest-frame 6 -- 10$\micron$ region, which can be complicated by a superposition of Polycyclic Aromatic Hydrocarbon (PAH) emission lines and silicate absorption features.  The peak of the cold dust emission in FIR luminous galaxies in the local universe occurs between 40$\micron$ to 150$\micron$ which corresponds to 120$\micron$ to 450$\micron$ at $z \sim$ 2.  In the next few years the $Herschel$ telescope will provide unprecedented amounts of data at these wavelengths; however, the recent flight of the BLAST telescope (Devlin et al. 2009) has already provided some of the first deep observations of distant galaxies at 250$\micron$, 350$\micron$ and 500$\micron$. 
\newline\indent
In this paper we take advantage of the BLAST observations of the Extended Chandra Deep Fields South field (ECDFS) to directly measure the FIR SEDs of $z \sim$ 2 galaxies.  Two star forming galaxies in the Kriek et al. (2008) NIR spectroscopic sample are located in that field and are detected or have upper limits at 24$\micron$, 70$\micron$, and 160$\micron$ in the FIDEL (Magnelli et al. 2009) and SWIRE (Lonsdale et al. 2003) surveys, and are also detected by BLAST (Dye et al. 2009).  The ECDFS was recently surveyed by the LABOCA instrument on the APEX telescope, and both galaxies were detected at 870$\micron$ (Weiss et al. 2009).  By combining these data sets we have produced some of the first FIR SEDs that sample the cool dust bump of $z \sim$ 2 galaxies with good resolution and allow us to constrain their L$_{IR}$ directly from the FIR SED.  Throughout this paper we assume a H$_{0}$ = 70 km s$^{-1}$ Mpc$^{-1}$, $\Omega_{m}$ = 0.3, $\Omega_{\Lambda}$ = 0.7 cosmology when computing luminosity distances.

\section{Galaxy Sample and Data}
The NIR spectroscopic sample of Kriek et al. (2008) is a magnitude-limited (K $\leq$ 19.7 Vega) sample of 36 galaxies with a median z $\sim$ 2.3.  The sample is dominated by massive galaxies (M $>$ 10$^{11}$M$_{\odot}$), approximately 50\% of which have detected emission lines.   
Two of the galaxies in the sample are extremely bright at 24$\micron$ (S$_{24\micron}$ $>$ 0.5 mJy), and both are located in the ECDFS (ID\#'s ECDFS-4511 and ECDFS-12514 from Kriek et al. 2008).   Both galaxies have strong H$\alpha$ emission and Kriek et al. (2008) used this to determine spectroscopic redshifts of $z = 2.122$ and $z = 2.024$.  The ID numbers of the galaxies in the BLAST survey of the ECDFS are BLAST J033138-274122 and BLAST J033242-275511.  They are also detected at 870$\micron$ in the 200-hr map from the LABOCA array on the APEX telescope (Weiss et al. 2009, ID\#'s LESS J033139.6-274120, and LESS J033243.3-275517).  
\newline\indent
Using the v0.5 public-release FIDEL\footnote{http://irsa.ipac.caltech.edu/data/SPITZER/FIDEL/} MIPS imaging data of the ECDFS we have measured total 24$\micron$ and 70$\micron$ fluxes for these galaxies.  For the 24$\micron$ data the fluxes were measured within 11$''$ diameter apertures using the SExtractor package (Bertin \& Arnouts 1996) and an aperture correction from the MIPS data handbook\footnote{http://ssc.spitzer.caltech.edu/mips/dh/} was applied.  For the 70$\micron$ data we used the 24$\micron$ centroid and measured fluxes within 32$''$ apertures and applied the aperture correction from the MIPS data handbook.  The ECDFS was observed at 160$\micron$ as part of the SWIRE survey (Lonsdale et al. 2003).  From the data we measured 160$\micron$ fluxes within 64$''$ apertures, again using the 24$\micron$ centroid.  The SWIRE 160$\micron$ data is much shallower than the FIDEL data; however, ECDFS-12514 is detected at the 1.5$\sigma$ level (a similar detection was made by Dye et al. 2009 using the SWIRE data).  ECDFS-4511 is not detected; however, we calculate a 2$\sigma$ upper limit based on empty aperture fluxes from the SWIRE photometry.
\newline\indent
In our SED fitting we use our measured 24$\micron$, 70$\micron$ and 160$\micron$ fluxes, the 250$\micron$, 350$\micron$, and 500$\micron$ fluxes measured by Dye et al. (2009), and the observed 870$\micron$ fluxes from Weiss et al. (2009).   For ECDFS-4511 we do not use the 160$\micron$ upper limit in the fitting.  These fluxes are listed in Table 1.
\newline\indent
A joint analysis of the FIDEL 24$\micron$ and BLAST FIR data by Chary \& Pope (2010) has shown that source blending is a problem for many of the BLAST detections.  We have used the 24$\micron$ data to test for contamination of the longer wavelength data of these galaxies by blended sources.  
\newline\indent
The galaxy ECDFS-12514 is well-isolated on the 24$\micron$ image.  The nearest 24$\micron$ source is 10$"$ away; however, it is extremely faint, only 4\% of the flux of the main galaxy.  The only nearby bright 24$\micron$ source is 21$"$ away which means if it was luminous at any of the longer wavelengths, excluding 500$\micron$, it should be resolved as a separate source.  That galaxy is only 15\% of the main galaxy flux and shows no evidence of a detection in any of the longer wavelength channels, suggesting that the 70$\rightarrow$870$\micron$ measurements for ECDFS-12514 are probably not contaminated by blended sources.  
\newline\indent
Galaxy ECDFS-4511 is not completely isolated.  There are two well-detected 24$\micron$ sources within 30$"$.  The closest is 7$"$ away and is 15\% of the flux of the main galaxy.  The next closest is 11$"$ away and is 37\% of the flux of the main galaxy.  If these companions are luminous in the FIR, neither would be resolved in the FIR data and could cause the FIR flux to be overestimated.  Given that at $z \sim$ 2, the SED of the main galaxy has been redshifted to where the dust SED peaks in the BLAST bands and the observed 24$\micron$/FIR flux ratio is maximal, we expect that the worst-case scenario would be if the nearby galaxies were at the same redshift and had similar FIR SEDs as the main galaxy and therefore contribute equally to the FIR flux.  If so, it suggests the FIR fluxes and L$_{IR}$ could be overestimated by as much as $\sim$ 30\%.  If the companions are at either higher or lower redshift, they are likely to have higher 24$\micron$/FIR flux ratios and therefore the overestimate is likely to be less.  As we discuss below, this possible overestimate of the L$_{IR}$ of ECDFS-4511 does not change any of the conclusions of this paper.
\begin{figure*}
\plotone{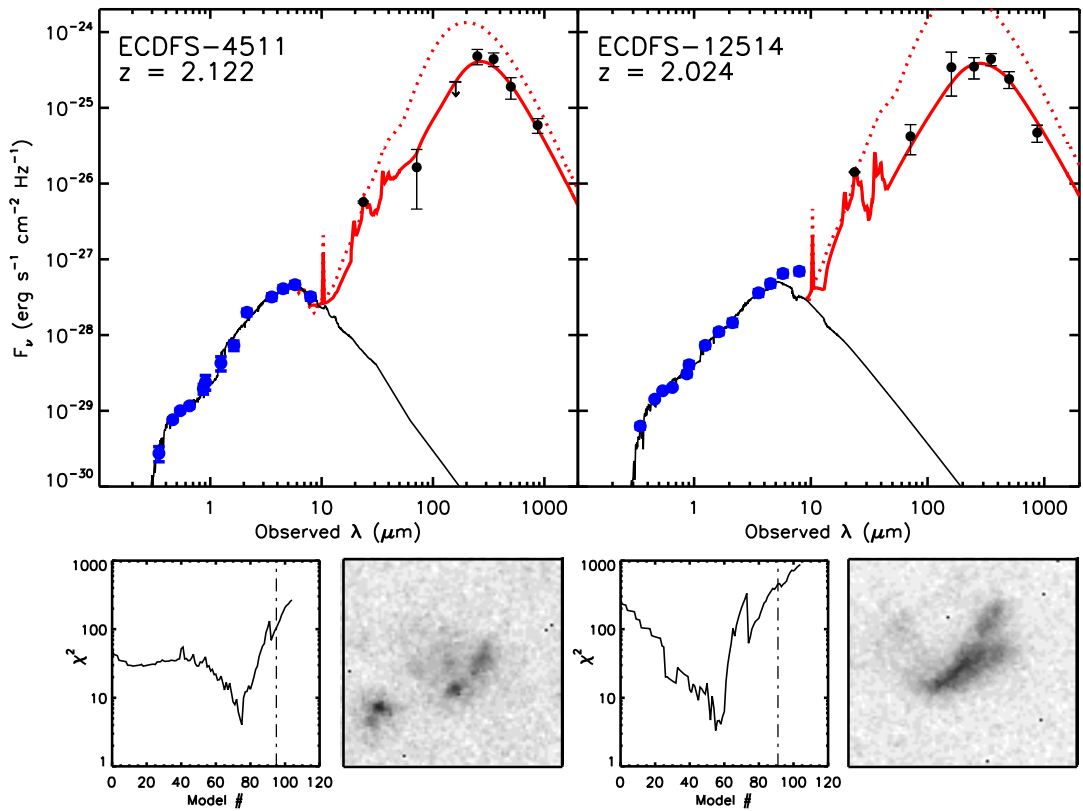}
\caption{\footnotesize Top panels: U $\rightarrow$ 870$\micron$ SEDs for galaxies ECDFS-4511 (left) and ECDFS-12514 (right).  The U$\rightarrow$8$\micron$ photometry is plotted in blue, and the best fitting BC03 model to those data is plotted in black.  The FIR photometry from Spitzer, BLAST and APEX are plotted in black.  The best fitting CE01 model to those data is shown in red.  The $\chi^2$ grid for the FIR SEDs is shown in the bottom left panels.  The dot dashed line in the upper panel shows the $\chi^2$ of the local template with the same L$_{IR}$ as the best-fit template.  Both galaxies are inconsistent with the local template and are best described as "scaled up" versions of lower luminosity galaxies.  The dotted FIR SED in the main panel shows the CE01 template that would be chosen based on only the observed 24$\micron$ flux.  For these galaxies those templates overestimate the L$_{IR}$ by factors of 3.0 and 6.2, respectively.  The bottom right panels show 3$"$ $\times$ 3$"$ NICMOS images of the galaxies.}
\end{figure*}
\begin{deluxetable*}{cccl}
\tabletypesize{\footnotesize}
\scriptsize
\tablecolumns{4}
\tablecaption{MIR/FIR Photometric Data}
\tablehead{\colhead{Wavelength} & \colhead{ECDFS-4511} & \colhead{ ECDFS-12514 } & \colhead{ Survey/reference }
  \\
\colhead{} & \colhead{flux (mJy)} & \colhead{flux (mJy)} & \colhead{}
}
\startdata
24 $\micron$ & 0.57 $\pm$ 0.02 & 1.42 $\pm$ 0.03 & FIDEL, this paper\\
70 $\micron$ & 1.6 $\pm$ 1.2 & 4.2 $\pm$ 1.8 & FIDEL, this paper\\
160 $\micron$ & $<$ 40 & 34 $\pm$ 20 & SWIRE, this paper\\
250 $\micron$ & 48 $\pm$ 11 & 35 $\pm$ 11 & BLAST, Dye et al. (2009)\\
350 $\micron$ & 44 $\pm$ 9 & 44 $\pm$ 8 & BLAST, Dye et al. (2009)\\
500 $\micron$ & 19 $\pm$ 6 & 24 $\pm$ 6 & BLAST, Dye et al. (2009)\\
870 $\micron$ & 5.9 $\pm$ 1.3 & 4.7 $\pm$ 1.2 & LESS, Weiss et al. (2009)\\


\enddata
\tablecomments{1) All fluxes are measured fluxes and do not include statistical deboosting corrections.  2) ECDFS-4511 has two 24$\micron$ sources
within 11$"$ of the main galaxy.  These galaxies cannot be resolved in the FIR data.  We estimate that if they have substantial FIR flux they could inflate the
FIR fluxes, and L$_{IR}$ by as much as 30\% (see text).}
\end{deluxetable*}

\section{Total Infrared Luminosities and Star Formation Rates}
\subsection{Spectral Energy Distributions and Fitting}
In the top panels of Figure 1 we plot the observed U $\rightarrow$ 870$\micron$ flux densities of the galaxies.  The 24$\micron$ $\rightarrow$ 870$\micron$ flux densities, which are from dust emission are plotted as black points and the U$\rightarrow$8$\micron$ flux densities, which are from stellar photospheres are plotted as blue points.  For reference, the best fit Bruzual \& Charlot (2003, hereafter BC03) model to the U$\rightarrow$8$\micron$ data from Muzzin et al. (2009) has also been plotted as the solid black line.  Figure 1 shows that there is an unambiguous peak in the FIR SEDs of the galaxies at $\sim$ 300$\micron$, a clear signature of the redshifted thermal emission from dust.
\newline\indent
We fit the 24$\micron$ $\rightarrow$ 870$\micron$ data using the models of Chary \& Elbaz (2001; hereafter CE01).  The CE01 models are a grid of 105 synthetic SEDs that were constructed to reproduce the correlations between the observed MIR and FIR fluxes of local galaxies.  Each model SED is associated with an L$_{IR}$ ranging between 3 $\times$ 10$^{8}$ L$_{\odot}$ and 6$\times$10$^{13}$ L$_{\odot}$.  We fit the 24$\micron$ $\rightarrow$ 870$\micron$ data using each of the 105 template SEDs using a $\chi^2$ minimization method where the amplitude of the SED is the only free parameter.  
\newline\indent
The galaxy ECDFS-4511 is the most FIR-luminous of the two galaxies and is well-fit by the SED template of a ULIRG.  In the CE01 models that template is associated with an L$_{IR}$ = 1.1$\times$10$^{12}$ L$_{\odot}$; however, ECDFS-4511 has L$_{IR}$ = 8.9$^{+1.1}_{-1.4}$$\times$10$^{12}$ L$_{\odot}$, eight times larger than the luminosity of the local template.  
\newline\indent
The second galaxy, ECDFS-12514, is three times brighter at 24$\micron$ than ECDFS-4511; however, it has a lower L$_{IR}$.  The best-fit template is a LIRG template with strong PAH features and is associated with L$_{IR}$ = 1.2$\times$10$^{11}$ L$_{\odot}$.  The L$_{IR}$ of ECDFS-12514 is 5.7$^{+1.2}_{-0.1}$$\times$10$^{12}$ L$_{\odot}$, almost 50 times more luminous than the local template.   If the local template is correct, the large 24$\micron$ flux from ECDFS-12514 likely arises from strong rest-frame 7.8$\micron$ PAH emission redshifted into the observed 24$\micron$ bandpass.
\newline\indent
The key result in Figure 1 is that although the $z =$ 2 galaxies have FIR SEDs that are similar to local FIR luminous galaxies, their luminosities are roughly an order of magnitude larger.  It appears that both galaxies are strongly "scaled up"  versions of much lower luminosity galaxies in the local universe.  The dot-dashed line on the $\chi^2$ grids in Figure 1 show the $\chi^2$ of the fit to the local template that is associated with the best fit L$_{IR}$ of ECDFS-4511 and ECDFS-12514.  Those $\chi^{2}$s are significantly worse than the best-fit $\chi^2$ and show that the SEDs of the $z \sim$ 2 galaxies are inconsistent with having the same SED shape of local templates of similar luminosity at $>$ 25$\sigma$.  
\subsection{Comparison of FIR and 24$\micron$ Star Formation Rates}
Direct constraints on the cool dust SED allow us to address the issue of how well L$_{IR}$, and SFR(L$_{IR}$) can be determined for $z \sim$ 2 galaxies using 24$\micron$ data alone.  The most common practice for determining L$_{IR}$ from 24$\micron$ photometry is to select the local template with an L$_{IR}$ that would match the observed 24$\micron$ flux if redshifted to the redshift of the source galaxy.  In Figure 1 we plot the CE01 template that would be selected based on using that method as the red dotted line.
\newline\indent
The extraordinary 24$\micron$ flux of both galaxies requires that they use the most luminous templates in the CE01 library.  The implied L$_{IR}$ for ECDFS-4511 and ECDFS-12514 is 2.8$\times$10$^{13}$ L$_{\odot}$ and 3.8$\times$10$^{13}$ L$_{\odot}$, respectively.  These templates overestimate the L$_{IR}$ from the full SED fit by factors of 3.0 and 6.2 respectively, and it is clear from Figure 1 that they cannot simultaneously fit both the 24$\micron$ flux densities and the FIR flux densities.  If these galaxies are representative of most FIR-luminous galaxies in the distant universe, it suggests we may be systematically overestimating total SFRs by similar factors.
\newline\indent
Similar results have also been found by Papovich et al. (2007) and Murphy et al. (2009). Papovich et al. (2007) compared L$_{IR}$ determined from 24$\micron$ data to that determined using the combined 24$\micron$, 70$\micron$, and 160$\micron$ photometry and found that for the most luminous galaxies (S$_{24}$ $>$ 250$\mu$Jy) the 24$\micron$ photometry overestimated the L$_{IR}$ by factors of 2 -- 10.  Using a sample of 14 MIR luminous galaxies at 1.4 $< z <$ 2.6 with Spitzer IRS spectra and SCUBA 850$\micron$ photometry Murphy et al. (2009) found that 24$\micron$ fluxes combined with local templates systematically overestimated the L$_{IR}$ by a factor of $\sim$ 5. 
\begin{figure}
\plotone{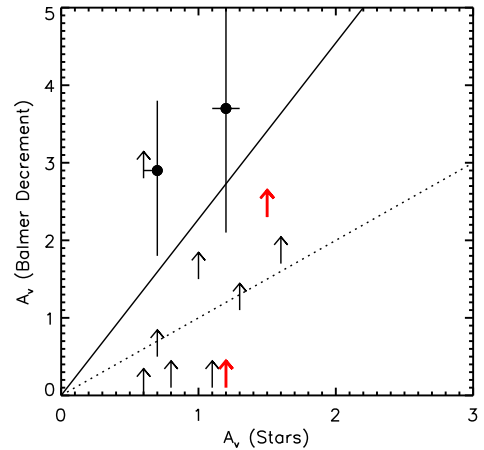}
\caption{\footnotesize A$_{v}$ determined from the Balmer decrement vs. A$_{v}$ determined from a fit to the galaxy stellar population for the 12 emission line galaxies from Kriek et al. (2007), Kriek et al. (2009a), and van Dokkum et al. (2005) with both H$\beta$ and H$\alpha$ within the NIR window.  Only two galaxies have secure H$\beta$ detections, the remainder are plotted as 1$\sigma$ lower limits.  The two BLAST-detected galaxies are indicated as the red lower limits.  The dotted line shows the one-to-one relation, and the solid line is the relation E(B-V)$_{stellar}$ = 0.44E(B-V)$_{nebular}$ suggested by Calzetti et al. (1997).  The two galaxies with H$\beta$ detections are in reasonable agreement with the Calzetti relation.  All of the galaxies with lower limits are consistent with the Calzetti relation but only about half are consistent with the one-to-one relation.}
\end{figure}
\subsection{Comparison of FIR and H$\alpha$ Star Formation Rates}
The FIR-derived SFRs can also be compared to the dust corrected H$\alpha$ SFRs for these galaxies.  H$\alpha$ equivalent widths (EWs) were measured for both galaxies using moderate-resolution NIR spectra from SINFONI by Kriek et al. (2007).  We convert their EW(H$\alpha$) to an H$\alpha$ line flux by multiplying by the continuum flux at 6563$\AA$ estimated using the BC03 SED model fits from Muzzin et al. (2009).  
\newline\indent
We assess the extinction for the H$\alpha$ flux using both the fit to the stellar SED (U$\rightarrow$8$\micron$) and the Balmer decrement.  For the SED fits we use the Muzzin et al. (2009) best-fit value of A$_{v}$ which is derived from fitting BC03 models using solar metallicity and the Calzetti et al. (2000) dust law (hereafter A$_{v,SED}$).  For the A$_{v}$ from the Balmer decrement (hereafter A$_{v,Balmer}$) we assume the Calzetti et al. (2000) dust law and an intrinsic ratio of H$\alpha$/H$\beta$ = 2.86.  Although both galaxies have high S/N H$\alpha$ detections in the SINFONI spectra, neither has a secure H$\beta$ detection.  Therefore instead of a true Balmer decrement we calculate 1$\sigma$ lower limits\footnote{Upper limits on H$\beta$ translate to lower limits on H$\alpha$/H$\beta$ and lower limits on A$_{v,Balmer}$} using the 1$\sigma$ upper limits on H$\beta$ determined by Kriek et al. (2007).  The H$\beta$ upper limits have been corrected for absorption by the stellar population using a BC03 SED fit to the broadband photometry and the NIR spectroscopy.
\newline\indent
In Figure 2 we plot A$_{v,Balmer}$ vs. the A$_{v,SED}$ for the two galaxies as the red lower limits.  For comparison, measurements of the 9 other $z \sim$ 2 galaxies in Kriek et al. (2007), and the galaxy from Kriek et al. (2009b) are also plotted.  Two of these galaxies (ID\#'s 1030-807 and CDFS-6202) have secure H$\beta$ detections, the remainder are plotted as 1$\sigma$ lower limits.  The one-to-one relation is shown as the dotted line.  
\newline\indent
The two galaxies with H$\beta$ detections do not agree well with the one-to-one relation.  The lower limits for approximately half the galaxies are also inconsistent with the one-to-one relation.  A similar disagreement amongst A$_{v}$ measurements was found for local galaxies by Calzetti (1997).  She showed that this difference can be understood as greater extinction towards the star forming regions, where the H$\alpha$ originates, than to the overall stellar population.  The Calzetti (1997) relation for local galaxies is E(B-V)$_{stars}$ = (0.44 $\pm$ 0.03)E(B-V)$_{nebular}$.  In Figure 2 we plot that relation converted to A$_{v}$ assuming R$_{v}$ = 4.05 (Calzetti et al. 2000) as the solid line.  The Calzetti (1997) relation seems to provide a better description of the overall population of $z \sim$ 2 star forming galaxies, albeit that most galaxies only have lower limits.  
\newline\indent
Given the uncertainty in the A$_{v,Balmer}$, and the well-determined A$_{v,SED}$ we have chosen to dust correct the H$\alpha$ flux using the A$_{v,SED}$ but using the E(B-V)$_{stars}$ = (0.44 $\pm$ 0.03)E(B-V)$_{nebular}$ relation.  We convert this to a SFR using the Kennicutt (1998) relation, SFR(H$_{\alpha}$) = 0.57 $\cdot$ L(H$\alpha$)/1.26$\times$10$^{-41}$ ergs s$^{-1}$, where the factor of 0.57 has been added to convert from a Salpeter IMF to a Chabrier IMF.  Applying these conversions we find that ECDFS-4511 and ECDFS-12514 have SFR(H$\alpha$) = 620$^{+260}_{-140}$ M$_{\odot}$ yr$^{-1}$ and 1210$^{+750}_{-40}$ M$_{\odot}$ yr$^{-1}$, respectively.  We note that the majority of the error budget for these measurements comes from the uncertainty in the extinction correction, not from random errors in the measurement of the H$\alpha$ line flux.
\newline\indent
We convert the L$_{IR}$ to a SFR using the Kennicutt (1998) relation, SFR(L$_{IR}$) = 0.57 $\cdot$ L$_{IR}$/5.8$\times$10$^{9}$L$_{\odot}$, where again the factor of 0.57 is to convert to a Chabrier IMF.  We derive  SFR(L$_{IR}$) = 870$^{+150}_{-140}$ M$_{\odot}$ yr$^{-1}$ and 560$^{+120}_{-10}$ M$_{\odot}$ yr$^{-1}$ for ECDFS-4511 and ECDFS-12514, respectively.  
\newline\indent
Comparing the two SFR indicators shows that they agree to within roughly a factor of 2 for both galaxies.  Such an agreement is good considering that there are still systematic uncertainties in the conversion of both L$_{IR}$ and L(H$\alpha$) to a SFR that are of that order (Kennicutt et al. 2009).  
\newline\indent
It is worth noting that the agreement in SFRs only occurs because L$_{IR}$ is determined from the full FIR SED fits, and because the extra extinction correction has been applied to the L(H$\alpha$).  If we do not use the L$_{IR}$ from the FIR SED fit and instead use the SFR(24$\micron$) we find that the IR SFR is larger than the SFR(H$\alpha$) by factors of 4.8 and 3.4 for ECDFS-4511 and ECDFS-12514, respectively.  Likewise, if we extinction correct the L(H$\alpha$) using the A$_{v,SED}$ without extra extinction we find that the SFR(L$_{IR}$) is larger than the SFR(H$\alpha$) by a factor of 5.4 and 2.5; however, this time because the SFR(H$\alpha$) is underestimated.   If we compare the SFR(24$\micron$) to the SFR(H$\alpha$)$_{SED}$, the SFR(24$\micron$) is larger by factors of 18.7, and 18.6.
\begin{figure}
\epsscale{0.9}
\plotone{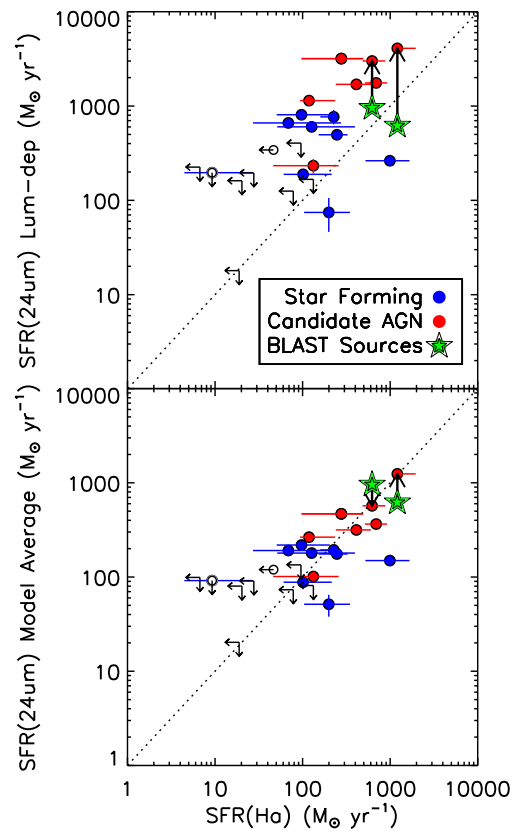}
\caption{\footnotesize Top Panel: SFR derived from the 24$\micron$ fluxes and the luminosity dependent CE01 templates vs. the SFR determined from the dust corrected L(H$\alpha$) for the full sample of K-selected galaxies in Kriek et al. (2008).  The luminosity dependent templates systematically overestimate the SFR(H$\alpha$) for the majority of galaxies.  Bottom Panel:  Log-average of the SFR derived from the 24$\micron$ fluxes and all 105 CEO1 templates vs. SFR(H$\alpha$).  An average of the local templates provides IR SFRs that are in better agreement with the SFR(H$\alpha$).  The luminosity dependent method requires that the majority of galaxies use the most luminous templates, whereas the template averaging forces the L$_{IR}$ to come from cooler templates.  The better agreement from cooler templates suggests the entire sample of massive $z \sim$ 2 galaxies would be consistent with being "scaled up" versions of lower luminosity galaxies. }
\end{figure}

\section{Are the BLAST-detected Galaxies Typical of \lowercase{z} $\sim$ 2 Massive Galaxies?}
The reasonable agreement between the FIR-derived and H$\alpha$ derived SFRs in the previous section shows that provided sufficient data to robustly determine both L$_{IR}$ and a dust corrected L(H$\alpha$), the locally-calibrated conversions of L$_{IR}$ $\rightarrow$ SFR and L(H$\alpha$) $\rightarrow$ SFR appear to provide consistent results for these galaxies.  Before we use these two galaxies to draw conclusions about the star formation properties of the larger population of $z \sim$ 2 massive galaxies, we must consider if they are representative.  With 24$\micron$ flux densities of 570 $\pm$ 18 $\mu$Jy and 1417 $\pm$ 28 $\mu$Jy, these two galaxies are the brightest 24$\micron$ sources in the entire Kriek et al. (2008) sample.  That they are the only ones detected in both the BLAST and APEX observations of the ECDFS is potentially a strong selection effect.
\newline\indent
One way to test how representative the BLAST-detected galaxies are of the entire K-selected sample is to compare the SFR(24$\micron$) and SFR(H$\alpha$) of the full sample and see if it shows the same disagreement as the BLAST-detected galaxies.  MIPS 24$\micron$ observations of the MUSYC fields were obtained as part of Spitzer program GO-30873 (PI: Labb\'{e}).  We reduced that data with the standard MOPEX procedures and derived 24$\micron$ fluxes for galaxies in the Kriek et al. (2008) sample using the method described in $\S$ 2.  There are 24 galaxies in the Kriek et al. sample where H$\alpha$ is observable in the NIR window from the ground, and of these galaxies 15 are detected at 24$\micron$.  The MUSYC 24$\micron$ data is much shallower than the FIDEL data, so the non-detection of some galaxies, depending on their SFR and redshift is expected.  For the undetected galaxies we calculate 5$\sigma$ upper limits based on empty aperture fluxes in the 24$\micron$ mosaics.
\newline\indent
In the top panel of Figure 3, we plot the SFR(24$\micron$) vs. the SFR(H$\alpha$) for these galaxies.  Galaxies that are candidates for hosting an AGN from either an X-ray detection, emission-line ratios (Kriek et al. 2007) or for having a powerlaw SED in the IRAC bands (Muzzin et al. 2009) are plotted as red symbols (this includes both BLAST-detected galaxies, see $\S$ 5.3).  For reference, the BLAST-detected galaxies are plotted as stars at their SED-fit SFR(L$_{IR}$) locations with arrows connecting them to their SFR(24$\micron)$.  Just like the BLAST sources, the majority of galaxies have a SFR(24$\micron$) that is systematically larger than the SFR(H$\alpha$).  This suggests that the local luminosity dependent templates probably overestimate the IR SFR of the majority of $z \sim$ 2 massive galaxies.
\newline\indent
Franx et al. (2008) and Wuyts et al. (2009) also noticed that the SFRs of $z \sim$ 2 galaxies tended to be larger when inferred from 24$\micron$ flux and the local luminosity-dependent templates than the SFR determined using a fit to the stellar SED (SFR(SED)).  They advocated using the L$_{IR}$ taken from the log-average of the local templates and argued that this produces SFRs that are in better-agreement with the SFR(SED).  For comparison, we determine the SFR(24$\micron$) using that method (although we use the CE01 templates, whereas they use the Dale \& Helou 2002 templates), and compare with the SFR(H$\alpha)$ in the bottom panel of Fig 3.
\newline\indent
Clearly the agreement between SFRs is much better when SFR(24$\micron$) is determined using template averaging method.  Given that when using the luminosity dependent templates, most of our galaxies require the most luminous templates to match the 24$\micron$ flux, it suggests the better agreement between SFRs with the template averaging method is because it includes numerous lower luminosity templates in the average.  This suggests that the larger sample of K-selected galaxies, like the BLAST-detected galaxies, can be described as scaled up versions of lower luminosity galaxies.  Furthermore, it appears that despite their exceptionally large FIR luminosities, the cooler SEDs of the BLAST-detected galaxies may be  representative of the larger population of massive galaxies at $z \sim$ 2.
\section{Discussion}
\subsection{Dust Temperatures and Evidence for "Scaled Up" Cool Galaxies}
It is remarkable that despite an L$_{IR}$ $\sim$ 10$^{13}$ L$_{\odot}$, both of the BLAST-detected $z \sim$ 2 galaxies are better fit with local templates that have PAH features and a cool "cold" dust bump.  In the nearby universe, galaxies with L$_{IR}$ $\sim$ 10$^{13}$ L$_{\odot}$ tend to have both the warmest "cold" dust (T$_{dust}$ $\sim$ 10 -- 70K, emission peaking in the FIR) and significant amounts of "hot" dust (T$_{dust}$ $\sim$ 70 -- 500K, emission peaking in the MIR).  Although both of these dust components can come from deeply embedded star formation (e.g., Tran et al. 2001), it is thought that the dominant radiation source in most, if not all nearby galaxies with L $\sim$ 10$^{13}$ L$_{\odot}$ is an AGN (e.g., Genzel et al. 1998; Lutz et al. 1998).  
\newline\indent
By contrast, the best fit templates of the $z \sim$ 2 galaxies are much cooler ULIRG and LIRG templates which are most frequently associated with star forming galaxies in the local universe (e.g., Armus 2009).  These templates also tend to have stronger PAH features that are more prevalent because the "hot" dust component from the AGN no longer dominates the MIR emission.
\newline\indent
The first claims that IR luminous galaxies at $z \sim$ 2 may be scaled up versions of lower luminosity cooler galaxies came from 24$\micron$ observations of submillimeter-selected galaxies (SMGs).  Pope et al. (2006) showed that the SEDs of SMGs peak at longer wavelengths than expected based on an extrapolation of their 24$\micron$ flux using local templates, arguing that SMGs are cooler at a given luminosity than the local templates.   Subsequent MIR spectroscopic observations of SMGs from $Spitzer$ by Lutz et al. (2005), Pope et al. (2008), and Menendez-Delmestre et al. (2009) all found strong PAH features and relatively weak MIR continuum in the SMGs suggesting much less "hot" dust than is usually found in local galaxies of comparable L$_{IR}$.  Similarly, Papovich et al. (2009) found a weaker than expected rest-frame 24$\micron$ luminosity for a lensed SMG, arguing for the lack of a "hot" dust component, and hence a scaled up cool galaxy.  
\newline\indent
With access to the FIR we can determine a dust temperature (T$_{d}$) independent of the MIR SED and compare this to the local T$_{d}$-L$_{IR}$ relation.  We fit the 160$\micron$ $\rightarrow$ 870$\micron$ data (50$\micron$ $\rightarrow$ 300$\micron$ rest-frame) to modified blackbody curves of the form S($\nu$) = A$\nu^{\beta}$B($\nu$,T$_{d}$), where A is a normalization, B is the Planck function, and $\beta$ accounts for frequency-dependent dust emissivity.  Even with detections in numerous bands the S/N of the data is too low to constrain both $\beta$ and T$_{d}$ simultaneously so we have assumed $\beta$ = 1.5, a typical value for submillimeter galaxies.  The 70$\micron$ points are omitted in the fitting because they correspond to rest-frame 23$\micron$ and are too far into the MIR to include in a single-component blackbody fit.  We note that including the 70$\micron$ data actually produces identical best-fit temperatures; however, the $\chi^2$ of fits are substantially larger because no single-temperature model can describe the 23$\micron$ $\rightarrow$ 300$\micron$ rest-frame simultaneously.    
\newline\indent
We find dust temperatures of 40$^{+2}_{-1}$K and 41$^{+5}_{-7}$K for ECDFS-4511 and ECDFS-12514, respectively.  These correspond to rest-frame monochromatic IRAS color measurements, C $\equiv$ log$_{10}$(f$_{60\micron}$/f$_{100\micron}$), of -0.05$^{+0.05}_{-0.03}$ and -0.03$^{+0.09}_{-0.20}$, respectively.  Comparing these to the C-L$_{IR}$ relation measured by Chapin et al. (2009) in the nearby universe using IRAS shows that these galaxies have temperatures typical of nearby galaxies with L$_{IR}$ = 10$^{11}$ -- 10$^{12}$ L$_{\odot}$.  They are also similar to the T$_{d}$ of local LIRGS/ULIRGS measured by Clements et al. (2010).  Therefore, with the combined Spitzer, BLAST and APEX data we can now confirm that some massive galaxies at $z \sim$ 2 not only have PAH-dominated MIR SEDs similar to lower luminosity local galaxies, but they also have "cold" dust SEDs consistent with being scaled up versions of lower luminosity galaxies.  This suggests that star formation may be occurring in a different environment within galaxies at $z \sim$ 2 compared to $z \sim$ 0.  
\newline\indent
Local galaxies with SFRs $>$ 100 M$_{\odot}$ yr$^{-1}$ tend to have compact nuclear starbursts which are the cause of the warmer dust temperatures (e.g., Sanders \& Mirabel 1996).  The $z \sim$ 2 galaxies have SFRs $\sim$ 1000 M$_{\odot}$ yr$^{-1}$, but similar dust temperatures which suggests that the star formation may be occurring in multiple large starburst regions throughout the galaxy, rather than concentrated in a single compact starburst which would presumably have a substantially hotter dust temperature.  Indeed, a lensed $z > 2$ galaxy with precisely such characteristics has recently been observed by Swinbank et al. (2010a) using interferometric submillimeter observations. 
\newline\indent
Kriek et al. (2009a) presented SINFONI-IFU H$\alpha$ spectra of both these galaxies.  Their maps show that the H$\alpha$ emission is resolved on scales of a few kpc, and is also consistent with the extended star formation hypothesis.  The greater availability of gas at high redshift may permit more spatially extended star formation at a high rate, something that is uncommon in local galaxies.  
\subsection{Morphologies}
In the inset of Figure 1 we show 3$"$$\times$3$"$ NIC2 F160W images of the galaxies from Kriek et al. (2009a).  At $z \sim$ 2 these correspond to rest-frame V-band morphologies and span an angular size of $\sim$ 25 kpc.  Like most of the high redshift ULIRGs and HLIRGs previously observed with HST (e.g., Chapman et al. 2003; Dasyra et al. 2008;  Swinbank et al. 2010b) the galaxies clearly have disturbed morphologies.  Kriek et al. (2009a) measure effective radii (R$_{e}$) of 5 and 3 kpc for ECDFS-4511 and ECDFS-12514, respectively, which means that both galaxies are comparable in size to local galaxies of similar mass.  Rest-frame optical morphologies make it difficult to identify the location of the star formation (which is presumably highly obscured); however, the disturbed nature of both galaxies shows that these are complex systems, possibly "stream-fed" galaxies (Dekel et al. 2009) or mergers.  Unlike the FIR SEDs, the morphologies do not appear to be scaled up versions of local star forming disk galaxies.   
\subsection{AGN Contribution to the MIR and FIR Emission}
Despite SEDs that are associated with local star forming galaxies, part of the MIR and FIR emission in both galaxies almost certainly comes from an AGN.  Using emission line ratios and the BPT diagram (Baldwin et al. 1981), Kriek et al. (2007) found that both galaxies were in the region of the diagram that Kewley et al. (2006) considered to have "composite" spectra with contributions from both star formation and an AGN.  
\newline\indent
ECDFS-4511 is detected in the 2 Ms $Chandra$ map of the CDFS (Luo et al. 2008), and has an L$_{x}$ $\sim$ 2$\times$10$^{44}$ erg s$^{-1}$, implying it hosts a Seyfert-luminosity AGN.  ECDFS-12514 is located in the 250 ks map of the ECDFS (Virani et al. 2006), but is undetected.   Although undetected in the X-rays, ECDFS-12514 is formally classified as an IRAC powerlaw galaxy (F$_{\nu}$ $\propto$ $\nu^{\alpha}$, $\alpha$ $<$ -0.5; e.g., Donley et al. 2007, see Figure 1) which makes it a candidate for an obscured AGN.  
\newline\indent
Without more data we cannot constrain the precise fraction of the MIR and FIR light that comes from the AGN; however, the consistency of the MIR/FIR SED with local templates that are star formation dominated, and the "composite" emission line ratios suggest that young stars are probably the dominant energy source heating the dust.   Pope et al. (2008) and Menendez-Delmestre et al. (2009) showed that the MIR SEDs of most SMGs (the BLAST-detected galaxies are formally SMGs) are still dominated by PAHs and that AGN only contribute $\sim$ 30\% of the total MIR emission.  Putting the evidence together the most plausible scenario is that both galaxies host a moderate-luminosity AGN, but the AGN does not dominate the MIR/FIR SED.  Nevertheless, we note that both our SFR(L$_{IR}$) and SFR(H$\alpha$) should be formally considered upper limits as the AGN will contribute to both of these.

\subsection{Implications for SFRs at $z \sim$ 2 and MIR-excess Galaxies}
The results of our FIR SED fitting show that caution is required when determining and interpreting the SFRs of $z \sim$ 2 galaxies based on limited data.  Using only 24$\micron$ photometry to infer SFRs at $z \sim$ 2 may work well for lower luminosity, or less obscured galaxies (e.g., Papovich et al. 2007; Reddy et al. 2010); however, the local templates can overestimate the SFR(L$_{IR}$) of the most FIR luminous galaxies by as much as a factor of $\sim$ 6  (see also Papovich et al. 2007; Murphy et al. 2009).  
Likewise, the Balmer decrement measurements for these galaxies seem to indicate extra extinction toward the star forming regions.  It appears that on average this relation is similar to that suggested by Calzetti et al. (1997); however, the lack of secure  H$\beta$ detections in our sample precludes quantitative statements.  To correctly determine the SFR(H$\alpha$) of individual galaxies, rather than ensemble averages requires high quality Balmer decrement measurements.  
\newline\indent
The extra extinction towards the star forming regions suggested by the Balmer decrements also indicates that the UV SFRs for some of these IR luminous galaxies may not be reliable because the A$_{v}$ is large enough that parts of the star forming regions are optically-thick.  The SFRs estimated from a fit to the stellar populations (SFR(SED); effectively a dust corrected UV SFR) by Muzzin et al. (2009) for ECDFS-4511 and ECDFS-12514 are 168$^{+41}_{-63}$ M$_{\odot}$ yr$^{-1}$ and 145$^{+160}_{-0}$ M$_{\odot}$ yr$^{-1}$, and underestimate the SFR(L$_{IR}$) of the galaxies by factors of 5.2 and 3.5, respectively.  If we compare the SFR(SED) to the SFR(24$\micron$) as in e.g., Daddi et al. (2007b) we find that these galaxies have MIR SFRs 18 -- 25 times greater than the UV SFRs, and are extremely strong "MIR excess" galaxies.  
\newline\indent
The purpose of the Daddi et al. (2007b) classification was to identify candidate obscured AGN, and given that it is likely that both galaxies contain an AGN it appears that method successfully identifies such sources.  However, MIR excesses of factors of 18 -- 25 would suggest that the AGN is the dominant MIR energy source for both galaxies, something that is unlikely given the shape of the MIR/FIR SEDs and the emission line ratios.  This calls into question whether galaxies with weaker inferred MIR excesses, such those near the cutoff of SFR(UV$_{corr}$)/SFR(MIR+UV) $\sim$ 3 suggested by Daddi et al. (2007b) have a significant contribution from an AGN.
\subsection{Comparison to Models}
Our results suggest that a substantial population of massive $z \sim$ 2 galaxies are "scaled up" versions of lower luminosity galaxies.  Clearly such a population needs to be understood within the context of modern galaxy formation and evolution models.  Recently, Hopkins et al. (2010) have used their merger-driven evolutionary model to predict the IR luminosity functions of galaxies and quasars between 0 $< z <$ 6.  In their model they argue that there is a threshold in L$_{IR}$ above which disks, merger-driven star formation, and obscured AGN are the dominant IR energy source of galaxies.  At $z = 0$ these limits are log(L$_{IR}$) $\geq$ 11.5 for merger-driven star formation, and log(L$_{IR}$) $\geq$ 12.5 for obscured AGN.  Based on the evolution of the merger rate and gas fraction to $z \sim$ 2 they suggest that these thresholds evolve in log(L$_{IR}$) to log(L$_{IR}$) $\geq$ 12.5 for merger-driven star formation and log(L$_{IR}$) $\geq$ 13.0 for obscured AGN.   With log(L$_{IR}$) $\sim$ 13.0, the BLAST galaxies lie at an L$_{IR}$ that is near the suggested transition point between merger-driven star formation dominated or AGN dominated. 
\newline\indent
Interestingly, the FIR SEDs of these galaxies do not look like those of the AGN-dominated HLIRGs detected by $ISO$ (Genzel et al. 1998), or of the average nearby ULIRG, which are mostly merger-driven starbursts (e.g., Sanders et al. 1998).  Instead, these galaxies look like scaled up versions of even lower luminosity systems, which are not universally associated with mergers or AGN.   Taken at face value, it appears they do not fit well within the Hopkins et al. (2010) description.  
\newline\indent
On the other hand, the NIC2 morphologies of the galaxies are clearly disturbed enough to be considered candidates for undergoing mergers (Kriek et al. 2009b).  If these galaxies are merger-driven starbursts, it may be that they are missing the centrally-concentrated starburst which produces the much hotter dust seen in these galaxies in the local universe.  As Hopkins et al. (2010) point out, the efficiency of a merger at funneling gas and stars into the core of the galaxy is proportional to the gas fraction, with higher gas fraction mergers being less efficient.  If both galaxies have high gas fractions most of the star formation could be occurring on much larger physical scales than similar mass starbursts in the local universe, hence the cool dust SED. 
\newline\indent
 Galaxies with similar stellar masses (M $\sim$ 10$^{11}$ M$_{\odot}$) and gas fractions as high as 80\% have been observed in the distant universe (e.g., Tacconi et al. 2010) and could be analogues of the BLAST-detected galaxies.  If that is the case, it begs the question of what are the descendants of extended, massive, gas-rich star forming galaxies at $z \sim$ 2?  Local spheroids seem like an obvious candidate; however, some fraction of the progenitors of those systems are clearly small and quiescent at $z \sim$ 2 (e.g., van Dokkum et al. 2010; Hopkins et al. 2010; Bezanson et al. 2009; van der Wel et al. 2008).  If the large and active BLAST-sources are also progenitors of massive local spheroids it suggests there may be very different evolutionary paths which arrive at the same galaxies.
\newline\indent
These galaxies also appear to be consistent with the "stream-fed" galaxies proposed by Dekel et al. (2009) and the scaled up "turbulent disks" discussed by Genzel et al. (2008).  In their model, Dekel et al. (2009) suggest that massive galaxies may grow quickly through rapid cold gas accretion and clumpy star formation which funnels stars into a central bulge.  Given that these cold flows often separate into multiple large star forming clumps, it might be expected that these clumps produce cooler dust temperatures at fixed SFR than centrally concentrated starbursts.  
\newline\indent
The SFRs of the BLAST galaxies are higher than the 100 -- 200 M$_{\odot}$ yr$^{-1}$ predicted for "stream-fed" galaxies by Dekel et al. (2009).  The BLAST galaxies are the most IR luminous of the K-selected sample of Kriek et al. (2008), so if they are "stream-fed" galaxies, they may just represent the most active tail of the distribution.
\newline\indent
High quality kinematic data could potentially distinguish between merger-dominated or accretion-dominated systems.  These galaxies do have SINFONI IFU observations; however, these data lack the S/N to reliably determine V/$\sigma$.  Therefore, at present based on the MIR/FIR properties and morphologies we cannot distinguish between evolutionary models such suggested by both Hopkins et al. (2010) and Dekel et al. (2009).  

\section{Summary}
Using a combination of data from the Spitzer, BLAST and APEX observatories we have presented some of the first well-sampled FIR SEDs for two $z \sim$ 2 galaxies with rest-frame optical spectroscopy.  In both galaxies we see a clear peak in the FIR emission at $\sim$ 300$\micron$ which is caused by the redshifted thermal emission from dust.  The FIR SEDs of the galaxies are well described by local templates; however, the L$_{IR}$ appears to be an order of magnitude larger than the L$_{IR}$ associated with the locally calibrated templates.  This suggests that these galaxies are likely to be scaled up versions of "cool" local galaxies, rather than analogues of nearby HLIRG templates.
\newline\indent
A comparison of SFR(L$_{IR}$) and SFR(H$\alpha$) for these two galaxies shows that they are consistent; however, if we use only the 24$\micron$ photometry and the local luminosity dependent templates to determine the SFR we find that this overestimates the SFR(H$\alpha$) by as much as a factor of 6.  A larger sample of 24 galaxies from the Kriek et al. GNIRS sample shows the same relation between SFR(24$\micron$) and SFR(H$\alpha$) and suggests that many massive galaxies at $z \sim$ 2 are likely to be scaled up versions of lower luminosity galaxies.  
\newline\indent
Although based on just a handful of galaxies, our results show that FIR luminous galaxies at $z \sim$ 2 are clearly different than their counterparts in the local universe.  In the coming years the $Herschel$ and JWST observatories will provide us with large amounts of high-quality FIR observations on the population of luminous galaxies at $z >\sim$ 2.  Likewise, upcoming NIR multi-object spectrograph will provide better H$\alpha$ and H$\beta$ measurements for those galaxies, and the ALMA observatory will provide spatially-resolved FIR maps of these galaxies.  This plethora of new, high-quality data will allow us to better understand how the population of luminous, "cool" star forming galaxies fit into the overall picture of galaxy evolution.

\acknowledgements

We thank the anonymous referee whose comments helped improve the quality of the paper.  Support for program HST-GO-11135.08 was provided by NASA through a grant from the Space Telescope Science Institute.

\acknowledgements

\end{document}